\documentclass{article}

\usepackage{arxiv}

\usepackage[cp1252]{inputenc}
\usepackage{textcomp}
\usepackage{amsmath}
\usepackage{amssymb}
\usepackage{graphicx}
\usepackage{array}
\usepackage[abs]{overpic}
\usepackage{color}
\usepackage{soul}
\usepackage{amsthm}
\usepackage{units}
\usepackage{pifont}
\usepackage{multirow}
\usepackage{stackengine}
\usepackage{hyperref}

\usepackage[FIGTOPCAP]{subfigure}

\usepackage{tikz}

\usetikzlibrary{angles,quotes}
\usepackage{gensymb}
\usepackage{pgfplots}
\usepgfplotslibrary{patchplots}
\pgfplotsset{compat=newest}

\graphicspath{{figures/}}
\usepackage{booktabs}       
\usepackage{amsfonts}       
\usepackage{nicefrac}       
\usepackage{microtype}      
\usepackage{lipsum}
\usepackage{graphicx}
\graphicspath{{figures/}}

\usepackage[square,numbers]{natbib}
\usepackage{textcomp}
\usepackage{amsmath}
\usepackage{amssymb}
\usepackage{graphicx, subfigure}
\usepackage{array}
\usepackage[abs]{overpic}
\usepackage{tabularx} 
\usepackage{multirow}
\usepackage{hyperref}

\makeatletter

\usepackage{color}
\usepackage{soul}
\usepackage{amsthm}

\usepackage{defcml_fonts}

\graphicspath{{figures/}}

\newcommand{\bs}{\boldsymbol}

\newcommand{\mdp}{\partial}
\newcommand{\mdiv}{\mathrm{div}_{\bs{x}}\,}
\newcommand{\mcurl}{\mathrm{curl}_{\bs{x}}\,}

\title{Machine Learning Aided Multiscale Magnetostatics}

\author{
 Fadi~Aldakheel\\
  Institute of Continuum Mechanics\\ Leibniz University Hannover, Germany \\
  \texttt{aldakheel@ikm.uni-hannover.de} \\ 
   \And
 Celal~Soyarslan\\
  Chair of Nonlinear Solid Mechanics \\ University of Twente, The Netherlands \\
  \texttt{c.soyarslan@utwente.nl} \\ 
   \And
 Hari~Subramani~Palanisamy\\
  Fraunhofer Innovation Platform \\ University of Twente, The Netherlands \\
  \texttt{h.subramanipalanisamy@utwente.nl} \\ 
   \And
 Elsayed~Saber~Elsayed\\
  Institute of Continuum Mechanics\\ Leibniz University Hannover, Germany \\
  \texttt{elsayed.stud.uni-hannover.de} \\ 
}

\begin{document}
\maketitle

\begin{abstract}
Computational material modeling using advanced numerical techniques speeds up the design process and reduces the costs of developing new engineering products. In the field of multiscale modeling, huge computation efforts are expected for modeling heterogeneous materials while trying to reach high accuracy levels. In this work, a machine learning approach, namely the convolutional neural network (CNN), is developed as a solution providing a high level of accuracy, while being computationally efficient. The input for the CNN model consists of two-/three-dimensional images of artificial periodic and biphasic microstructures in the form of nonoverlapping and overlapping, mono- and polydisperse circular/spherical disk systems, which are generated by a random sequential inhibition process. These correspond to Statistical Volume Elements (SVE). Considering linear magnetostatics at the microscale, the output is the apparent permeability of the SVE. Training and testing data for the apparent properties is produced with finite element method-based two-scale asymptotic homogenization. The model efficiency is revealed by employing some representative examples in two- and three-dimensional settings. In this regard, the performance of the CNN model is assessed with the applied computational homogenization method relating to the accuracy and computational efficiency. The results with the CNN model show high accuracy in predicting the homogenized permeability and a significant decrease in computation time.
\end{abstract}
%
\section{Introduction}
\label{sec1-CNN}
Engineering materials often have heterogeneous microstructures composed of constituents at various scales; see, e.g., \cite{BARGMANN2018322}. The properties of these constituents, whether geometrical or physical, play a crucial role in determining the material's macroscopic properties. Understanding the relationship between the microstructure and macroscopic physical properties is crucial for material design optimization and manufacturing control. In this regard, scientists revert to mathematical modeling approaches such as analytical bounds \citet{voigt1887,reuss29}, effective medium theories, e.g., the Maxwell, self-consistent, differential effective-medium approximations, e.g., \citet{hashin+shtrikman62a,hashin+shtrikman63}, or {hierarchical} \cite{Michel, Fish2014} and {concurrent} \cite{Lloberas, Fish2014, Fish3,aldakheel2021simulation,aldakheel2020global} multiscale computational homogenization techniques. The former analytical methods provide fast but usually inaccurate predictions as their formulation incorporates limited microstructural descriptors. While the latter numerical homogenization methods provide high-fidelity solutions, they require large computational times, making their use impossible in real-time control. This motivates the development of computationally feasible approaches in this realm. A recently emerging third alternative in literature to this end is employing data-driven surrogate models devising machine learning, see, e.g.,   \cite{Zhang2020homogenization,lu2021deepxde,vlassis2022component,fuchs2021dnn2,lopez2018manifold,heider2021multi,henkes2022physics,zohdi2022note,bock2021hybrid,zhang2022hidenn,fernandez2020application,as2022mechanics,benaimeche2022k,aldakheel2021feed,cueto2022thermodynamics,mortazavi2022first,zohdi2022machine,bessa2019bayesian,fuhg2022physics,stocker2022novel,javvaji2022machine,kalina2022fe} and the references therein. Deploying the training costs offline materializing simulation or experimental data, these models surpass conventional rule-based approaches by drastically reducing the computational cost required during the prediction phase \cite{bahmani2021kd,prume2023model,lu2021stochastic}.

Being among machine learning approaches, Convolutional Neural Networks (CNNs) are commonly used in image evaluation tasks such as image classification \cite{NIPS2012_c399862d}, object detection \cite{Redmonetal2015}, and image segmentation \cite{Longetal2014}. While accomplishing these tasks, CNNs process an input image in multiple layers, each of which learns to extract distinct features from the image. This unique ability to automatically extract increasingly complex features from imagery promises huge potential in exploring structure-property relations in the modeling of heterogeneous materials, see, e.g.,  \cite{rao2020three,yang2018deep,eidel2023deep}. With this motivation, the key goal of this contribution is to extend the application of CNNs to multiscale linear magnetostatics through the development of a computationally accurate and efficient solution architecture. An important aspect is the illustration of advanced transfer learning of the developed CNN model to cases involving microstructural variations, the details of which have not been considered so far in the aforementioned earlier works. We seek estimates of the apparent intrinsic magnetic permeability tensor components of two-phase random composites in the application problems.
While doing so, we highlight the remarkable efficiency of the proposed model by employing different microstructures and comparing the results with the classical homogenization approaches.

 The training/validation data are synthetically produced using finite element-based first-order computational homogenization through the use of fundamental micro-to-macro homogenization principles, e.g., \citet{hill72}, \citet{suquet87}, and \citet{nemat-nasser+hori99}.
 Here the microstructural images are treated as the training input and the effective magnetic permeability tensor as labels.
 The studied two-/three-dimensional microstructures are monodisperse and polydisperse circular/spherical inclusions embedded in a homogeneous matrix. A sequential inhibition process with periodicity constraint is used for the random inclusion distributions. For regular periodic microstructures, the effective material properties are determined using asymptotic homogenization applied over a repeating unit cell, see, e.g., \citet{bensoussan+lions+papanicolaou78} and \citet{sanchez-palencia80}, provided that scale separation exists. For stochastic microstructures, conventionally, a sufficiently large representative volume element (RVE) is used. According to \cite{HILL1963}, a sufficiently large size should produce the same effective properties independent from the applied boundary conditions. This usually leads to  large RVE sizes, especially for materials with highly contrasting constituent properties \cite{SAB1992,ShenBrinson2006}. An alternative method is to use an ensemble of smaller VE sizes, referred to as the statistical volume elements, and to compute the effective properties, apply averaging over this ensemble. The homogenized property of each ensemble member is then referred to as apparent rather than effective. Similar to \cite{rao2020three}, we use SVEs in our computations.

The paper is organized as follows:
In Section~\ref{sec2-CNN}, a brief overview of the micro-to-macro transition concept is introduced.
Next, the convolutional neural networks (CNNs) theory is presented in detail in Section~\ref{sec3-CNN}.
The CNN model is then employed to predict the homogenized macroscopic stress of a microstructure representing a heterogeneous composite in Section~\ref{sec4-CNN}.
The model capability is illustrated through various representative examples in Section~\ref{sec5-CNN} and compared with the traditional multiscale methods. The trained network is then applied to learn the constitutive behavior of magnetostatics materials within the finite element application. Thereafter, the trained model is used to predict the effective (macroscopic) permeability tensor, and through transfer learning, it is applied to a new structure. Section~\ref{sec6-CNN} presents a summary and outlook for extensions of  this work.
%
\section{Homogenization in Linear Magnetostatics}
\label{sec2-CNN}
%
%
Let $\mathcal{B}\subset\mathcal{R}^3$ denote the homogenized macro continuum, a typical point  $^\textrm{M}\boldsymbol{x}\in\,\mathcal{B}$ of which encapsulates a microstructure, represented by the unit cell  domain $\mathcal{V}\subset\mathcal{R}^3$ bounded by  $\partial \mathcal{V}$ consisting of two constituent material domains $\mathcal{V}_1\subset\mathcal{V}$ and $\mathcal{V}_2\subset\mathcal{V}$.  The coordinate $^\textrm{M}\boldsymbol{x}$ represents the macroscale position vector, also referred to as the global or slow variable. Then, we denote the microscale position vector, also referred to as the local or fast variable, by $\boldsymbol{x}$.
\subsection{Magnetostatics at Microscale}
Classical electromagnetic field theory is governed by Maxwell's equations, which consist of Faraday's law, Maxwell-Amp\`{e}re law, electrical and magnetic Gauss law, and finally, the equation of continuity, as given in Eqs.\
(\ref{E:maxwell_5}), respectively\footnote{Let $s(\bs{x},t)$ denote a scalar field and $\bs{C}(\bs{x},t)$ and $\bs{D}(\bs{x},t)$ two vector fields, distributed over the domain represented by material
points $\bs{x}$ at time $t$. Using Einstein's summation convention $\bs{C}=C_i\bs{e}_i$ and  $\bs{x}=x_i\bs{e}_i$ where $\bs{e}_i$ represents Cartesian basis vectors are $x_i$ associated vector components. Considering that $\bs{\nabla}_{\bs{x}}$ gives the gradient
operator, the gradient of a scalar field reads
\begin{equation*}
\mathrm{grad}_{\bs{x}}\,s:=\bs{\nabla}_{\bs{x}} s :=\dfrac{\partial s(\bs{x})}{\partial x_{i}} \boldsymbol{e}_{i}\,.
\end{equation*}
Considering that $\cdot$ and $\times$ denote single-contraction (scalar) and cross products with $\bs{C}\cdot\bs{D}=C_iD_i$ and $\bs{C}\times\bs{D}=\epsilon_{ijk}C_jD_k\bs{e}_i$ where $\epsilon_{ijk}$ is the Levi-Civita symbol, such that
$\epsilon_{ijk}$ is $1/-1$ for even/odd permutations of  $(i, j, k)$ and 0 for repeated indices, the curl and the divergence of a vector field $\bs{C}(\boldsymbol{x})$, respectively, read%
\begin{align*}
\mathrm{curl}_{\bs{x}}\,\bs{C}:=\bs{\nabla} \times \boldsymbol{C}:=
-\dfrac{\partial \bs{C}(\bs{x})}{\partial x_{i}}\times \bs{e}_{i}=
\epsilon_{kji}\frac{\partial C_{i}}{\partial x_{j}}\bs{e}_{k}\quad\text{and}\quad
\mathrm{div}_{\bs{x}}\,\bs{C}:=\bs{\nabla}\cdot \bs{C}:=\frac{\partial
\bs{C}(\bs{x}) }{\partial x_{i}} \cdot \bs{e}_{i}\,.
\end{align*}}  \cite{maxwell1873,jin2014}
\begin{align}
\mcurl\bs{E}=-\dfrac{\mdp{\bs{B}}}{\mdp t}\,,\quad
\mcurl\bs{H}=\dfrac{\mdp{\bs{D}}}{\mdp t}+\bs{J}\,,\quad
\mdiv\bs{D}=-\varphi\,,\quad
\mdiv\bs{B}=0\quad\text{and}\quad
\mdiv\bs{J}=-\dfrac{\mdp \varphi}{\mdp t}\,. \label{E:maxwell_5}
\end{align}
Here, $\varphi(\bs x,t)$ represents the scalar electric charge density and $\partial\{\bullet\}/\partial t$ the time rate of change of the term $\{\bullet\}$. $\bs E(\bs x,t)$, $\bs D(\bs x,t)$, $\bs H(\bs x,t)$, $\bs B(\bs x,t)$, $\bs J(\bs x,t)$ denote time- and position-dependent electromagnetic vector fields which are referred to as electric field intensity, electric flux density (also known as electric displacement or electric induction), magnetic field density, magnetic flux density (also known as magnetic induction) and electric current density, respectively. Only three of the five equations given in Eqs.\ (\ref{E:maxwell_5}) are independent. Limiting ourselves to  linear electromagnetics, and considering that $\bs \epsilon$, $\bs  \mu$ and $\bs  \sigma$ respectively denote second-order electrical permittivity, magnetic permeability and electrical conductivity tensors,
the constitutive relations between vector fields can be written as follows\footnote{Isotropy assumption allows representation of physical property tensors in terms of spherical tensors, hence scalars, such that  $\bs\epsilon=\epsilon\bs 1$, $\bs\mu=\mu\bs 1$ and $\bs\sigma=\sigma\bs 1$. As a consequence Eqs.\ (\ref{E:const_3}) can be represented as  $\bs{D}=\epsilon\bs{E}$, $\bs{B}=\mu\bs{H}$ and $\bs{J}=\sigma\bs{E}$, respectively. Here, considering that $\otimes$ denotes the dyadic product, $1=\delta_{ij}\bs e_i\otimes \bs e_j$ for $i,j=1,2,3$ with $\delta_{ij}$ denoting Kronecker's delta where $\delta_{ij}=1$ for $i=j$, and $\delta_{ij}=0$, otherwise.}
\begin{align}
\bs{D}=\bs\epsilon\cdot \bs{E}\,,\quad
\bs{B}=\bs\mu\cdot \bs{H}\quad\text{and}\quad
\bs{J}=\bs\sigma\cdot \bs{E}\,.   \label{E:const_3}
\end{align}
For static conditions, the vector fields do not depend on time $\varphi(\bs x,t)\rfloor \varphi(\bs x)$, $\bs E(\bs x,t)\rfloor \bs E(\bs x)$, $\bs D(\bs x,t)\rfloor  \bs D(\bs x)$, $\bs H(\bs x,t)\rfloor \bs  H(\bs x)$, $\bs B(\bs x,t)\rfloor \bs B(\bs x)$, $\bs J(\bs x,t)\rfloor \bs J(\bs x)$, thus, their time rate of change boils down to zero. The interaction between electric and magnetic fields does not hold, and one has decoupled electrostatic and electromagnetic problems. Moreover, if, for the magnetostatic case, the free current density is ignored, one has $\mcurl{\bs H }=\bs 0$. This implies that $\bs H$ is conservative, that is, it can be derived from a magnetic scalar potential (with the unit of Ampere), say $\varrho$, with $\bs H =-\mathrm{grad}_{\bs{x}}\,\varrho$, analogical to electrostatics in which the electric field $\bs E$ derived from an electric scalar potential $\phi$ with $\bs E=-\mathrm{grad}_{\bs{x}}\,\phi$. As a consequence, the conservation of magnetic induction and associated constitutive laws posed at the microscale are given as
\begin{align}
\mdiv\bs{B}=0\,,\quad
\bs{B}=\bs\mu\cdot \bs{H}\quad\text{and}\quad
\bs{H}=-\mathrm{grad}_{\bs{x}}\,\varrho\,. \label{E:micro_3}
\end{align}

\subsection{Problem at Macroscale}
\noindent The fields at the microscale are upscaled using the averaging operators through integration over the volume $\left\langle
\left[  \text{\ding{117}} \right]  \right\rangle $ and surface $\left\{ [ \text{\ding{117}}  ]  \right\}$ which are described as follows
\begin{align}
\left\langle [\text{\ding{117}} ] \right \rangle =\dfrac{1}{|\mathcal{V}|}
\int_{\mathcal{V}} \left[\text{\ding{117}} \right] \mathrm{d}V\quad\text{and}\quad
\left\{ [\text{\ding{117}} ]  \right\}  =\frac{1}{|\mathcal{V}|}\int_{\partial\mathcal{V}}\left[
\text{\ding{117}} \right] \mathrm{d}S\,.
\end{align}
For a concise notation, we use the overline notation which stands for volume averaging. Considering a generic tensor field $\bs{C}$ of any order we write $\overline{\bs{C}}=\left\langle \bs{C} \right \rangle$. The abovementioned averaging operators provide solutions to the macroscopic fields using the microscopic ones. While doing that, however, appropriate boundary conditions provide equivalence between the averaged-out (incremental) microscopic energies and the (incremental) macroscopic ones. This condition is also referred to as is referred to as Hill-Mandel condition. Considering that any micro-field $[\text{\ding{117}} ]$
can be additively decomposed into its volume averaged part, that is, the mean  $\overline{[\text{\ding{117}} ]}$ and the fluctuating part, which is
$[\text{\ding{117}} ]^{\prime}$, with $[\text{\ding{117}}]=\overline{[\text{\ding{117}} ]}+[\text{\ding{117}} ]^{\prime}$
for the case of linear magnetostatics Hill-Mandel condition satisfies the following equivalence\footnote{This can equivalently be stated with
$\overline{\bs{H}\cdot\bs{B}}=\overline{\boldsymbol{B}}\cdot\overline{\boldsymbol{H}}$.
}
\begin{equation}
\overline{\boldsymbol{B}^{\prime}\cdot \boldsymbol{H}^{\prime}}=
\overline{\boldsymbol{B}^{\prime}}\cdot\overline{\boldsymbol{H}^{\prime}}=0\,.
\label{E:hillmandel}
\end{equation}
Considering that $\bs H=\overline {\bs H} + \bs H'$, and supposing that the scalar field also includes a fluctuating part with $\varrho=\overline{\varrho}+\varrho'$ with $\overline{\varrho}=\overline{\bs H} \cdot \bs x$ and $\bs H'=\mathrm{grad}_{\bs{x}}\,\varrho'$, one finds the following boundary conditions which satisfy Hill-Mandel condition encapsulated in Eq.\ \eqref{E:hillmandel}, see, e.g., \cite{Chatzigeorgiou2014,ZABIHYAN2018105}.
\begin{align*}
1) \quad & \varrho  =\overline {\bs{H}}\cdot\boldsymbol{x}+\varrho' \text{ on }%
\partial \mathcal{V} \text{ with}\\
 & \text{a)}\quad \varrho'=0 \,,\\
 & \text{b)}\quad \varrho' \text{ periodic } t=\bs B\cdot \bs n \text{ antiperiodic}\,,\\
2) \quad & \boldsymbol{B}\cdot \boldsymbol{n} =\overline{\boldsymbol{B}}\cdot
\boldsymbol{n}\text{ on }\partial\mathcal{V}\,.%
\end{align*}
The boundary conditions given in 1(a), 1(b), and 2 are referred to as Uniform Dirichlet Boundary Conditions (UDBC), Periodic Boundary Conditions (PBC), and Uniform Neumann Boundary Conditions (UNBC). In the application problems of the current work, only PBC is used.
\subsection{Computation of Effective Permeability Tensor Components}
In our applications, we assume a generalized constitutive law for the macro continuum which reads $\overline{\bs B} = \bs \mu^\star\cdot \overline{\bs H}$ and $\overline{\bs H} = \bs \rho^\star\cdot \overline{\bs B}$ where $\bs \mu^\star$ and $\bs \rho^\star$ respectively denote the effective (macroscopic) permeability tensor and its inverse, with $[\bs \mu^\star]^{-1}=\bs \rho^\star$. We aim to find the components of these tensors. The effective permeability tensor  $\bs \mu^\star$ and its inverse have the following matrix representations
\begin{align}
\lfloor \bs \mu^\star \rfloor=
\begin{pmatrix}
\mu^\star_{11} & \mu^\star_{12} & \mu^\star_{13}\\
\mu^\star_{21} & \mu^\star_{22} & \mu^\star_{23}\\
\mu^\star_{31} & \mu^\star_{32} & \mu^\star_{33}
\end{pmatrix}\text{ and }
\begin{pmatrix}
\rho^\star_{11} & \rho^\star_{12} & \rho^\star_{13}\\
\rho^\star_{21} & \rho^\star_{22} & \rho^\star_{23}\\
\rho^\star_{31} & \rho^\star_{32} & \rho^\star_{33}
\end{pmatrix}\,.
\end{align}
Using PBC, the emerging magnetic induction field vectors $\overline{\bs B}_{(i)}$ for $i=1,2,3$ averaged over the representative volume element subjected to unit magnetic vector fields $\overline{\bs H}_{(i)}$  with $\lfloor\overline{\bs H}_{(1)}\rfloor=(1,0,0)^\top$, $\lfloor\overline{\bs H}_{(2)}\rfloor=(0,1,0)^\top$ and $\lfloor\overline{\bs H}_{(3)}\rfloor=(0,0,1)^\top$,  give  column $i$  of the permeability matrix, that is, $\lfloor\overline{\bs B}_{(1)}\rfloor\rightsquigarrow(\mu^\star_{11},\mu^\star_{21},\mu^\star_{31})^\top$, $\lfloor\overline{\bs B}_{(2)}\rfloor\rightsquigarrow(\mu^\star_{12},\mu^\star_{22},\mu^\star_{32})^\top$ and $\lfloor\overline{\bs B}_{(3)}\rfloor\rightsquigarrow(\mu^\star_{13},\mu^\star_{23},\mu^\star_{33})^\top$. A derivation of this result using  a first-order asymptotic periodic homogenization scheme is described in \cite{Soyarslan2022arxiv},  and \cite{SOYARSLAN2019103098} and the references therein for the case of magnetostatics and mechanics, respectively.

\section{Theory of Convolutional Neural Network (CNN)}
\label{sec3-CNN}

To overcome the issue of large computational costs to solve the multiscale problem on the micro-to-macro level, summarized in Section \ref{sec2-CNN}, an approach to replace the calculation with a machine learning model is introduced. Applying deep learning (DL) is an empirical, highly iterative process that requires training several models to reach satisfactory results. During this process, a combination of different parameters and hyper-parameters is tested.

This section aims to give a brief insight into the structure and operation of Convolutional Neural Networks (CNN) instead of going into specific variants and manifestations of this technique. CNN is a kind of Artificial Neural Network associated with deep learning techniques due to its structural configuration connecting multiple neuron layers to perform desired operations. More specifically, the CNN model is a specialized Feed Forward Neural Network (FFNN) that excels at working with data that have a grid-like structure, such as images or time-series data. CNN has success with computer vision applications, such as image classification, object detection, text detection, and recognition, see \citet{gu2018recent}. Its unique weight-sharing capabilities achieved by convolutional and pooling layers are the main differentiator between CNN and FFNN. Those two important operations and other deep learning functions will be explained next.

\begin{figure}[htb!]
	\centering
	\includegraphics[width=.65\columnwidth]{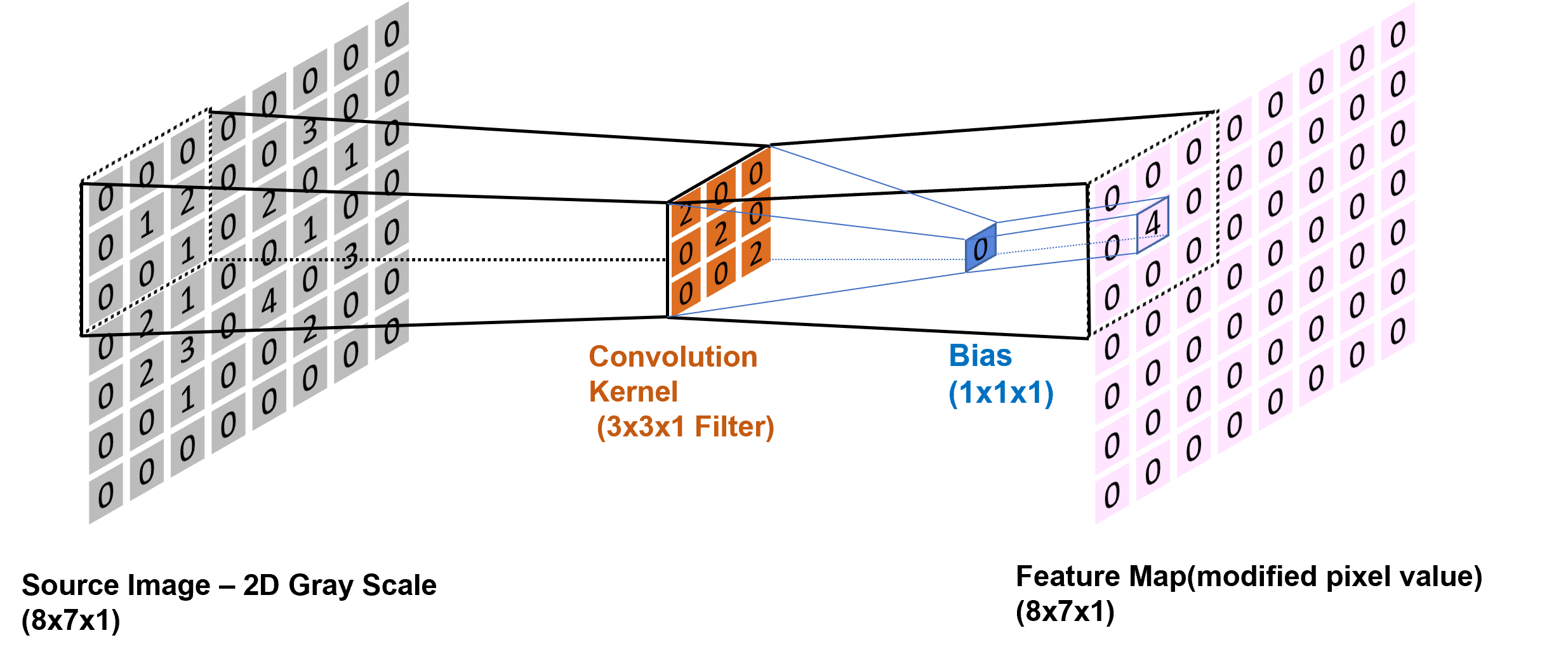}
	\caption{An example of a convolution operation added with bias performed on a grayscale image using a single convolution kernel to modify source pixel value.}
	\label{convolution_operation_gray}
\end{figure}

\begin{figure}[b]
	\centering
	\includegraphics[width=.65\columnwidth]{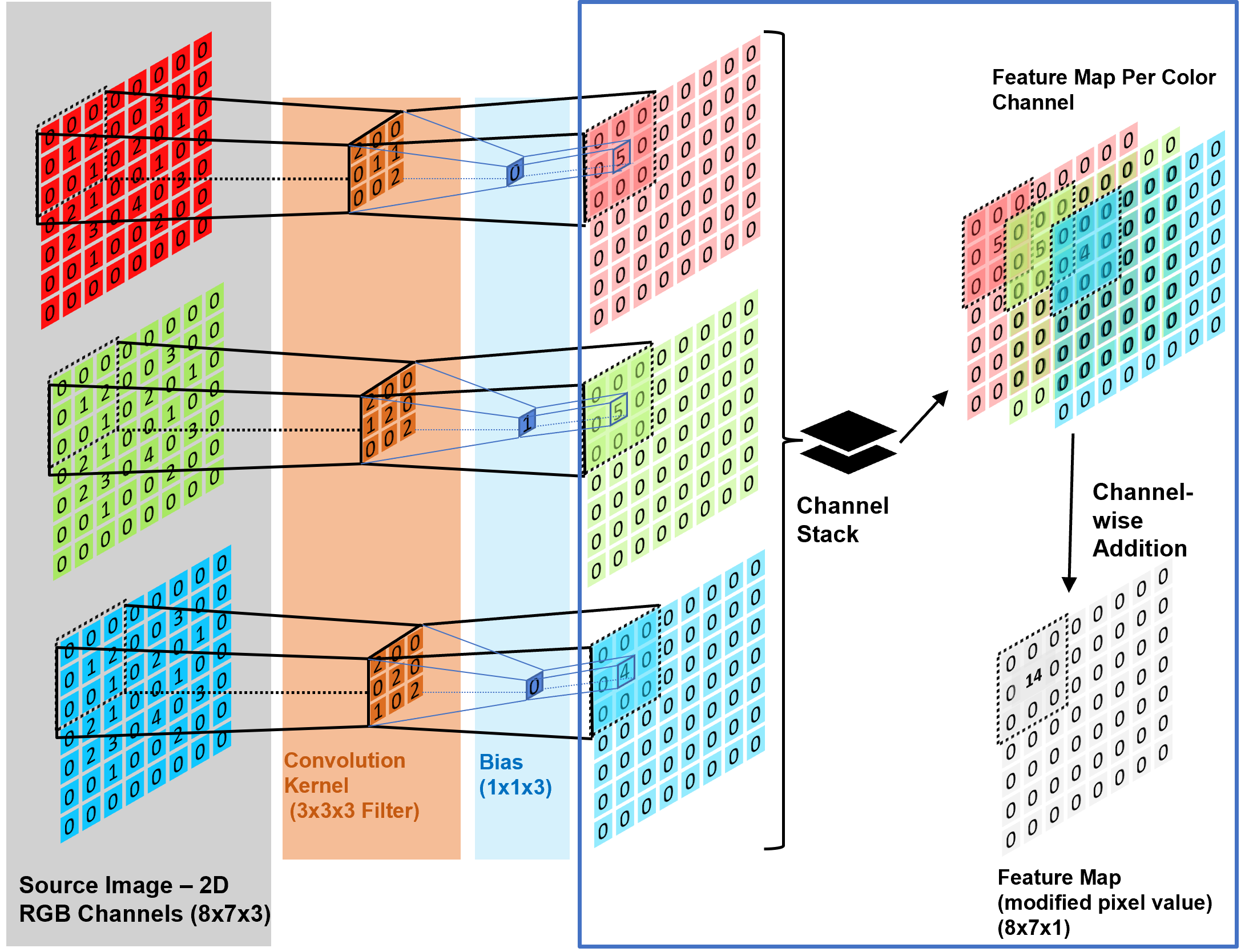}
	\caption{An example of an exploded view of a convolution operation added with bias performed on RGB channels using a single convolution kernel.}
	\label{convolution_operation_rgb}
\end{figure}

\subsection{Convolution Operation}
The convolution operation could be defined as a linear multiplication between an input and a set of weights. The input can be a gray-scale image or red-green-blue (RGB) scale image, and the set of weights is a matrix known as a \emph{filter} or \emph{kernel}. In this work, the input is an image with a dimension of  $d^{n_\mathrm{dim}}$ with $n_\mathrm{dim}=1,2,3$ for one-, two- and three-dimensional applications and pixel intensity values ranging between [0,1]. Where convolution operation is performed on localized receptive regions based on the chosen kernel dimension to generate a resulting modified pixel value. The sequential sliding window approach allows for manipulating pixels of an entire image. Performing convolution operations with a kernel at a different location in an image allows it to capture the same feature regardless of its spatial location and helps in learning the correlation between neighboring pixel values. The output of the convolutional operation is usually referred to as a \emph{feature map} or \emph{output image}. For a gray-scale image, the feature value at a position ($i,j$) in a feature map is given by the equation
\begin{equation}
z_{i,j} = \sum_m\sum_n \; \BI_{i+m,j+n} \; \Bk_{m,n}\; + \Bb_{1,1} ,
\end{equation}
where $\BI$ is the image, $\Bk$ is the applied filter with $m,n$ dimensions, and  $\Bb$ is the bias with a  dimension of $1\times1$. The convolution operation for the grayscale image is visualized in Fig \ref{convolution_operation_gray}. The detailed channel-based convolution operation for the RGB image with channel stacking is presented in Fig \ref{convolution_operation_rgb}.
The following equation can control the dimensions of the feature map for an input image
\begin{equation}
O_{od} = [I_{id} - F +2P] /S +1 ,
\end{equation}
where $\BO$ is the output feature map with $od$ square dimension, $\BI$ is the input image with $id$ square dimension, $\BF$ represents the size of square filter kernel, $\BS$ is the stride length for the sliding window operation, and $\BP$ is the number of zero padding applied to outer borders of an input image. Based on the architectural requirements, the output feature map dimension can be the same as an input image by assigning padding value to $P=[F-1]/2$.
As the convolution operation results in linear mapping between the given input and corresponding output, they fail to capture the complex feature relationship that can be non-linear. Therefore non-linearity is introduced to each feature value in the feature map element-wise with the help of an activation function

\begin{equation}
a_{i,j} = f(z_{i,j}) \; ,
\end{equation}
Note that the activation functions, also known as transfer functions, greatly impact the performance of deep learning models. In the early works of CNN, Sigmoid activation functions were used due to its simple yet effective non-linear mapping. However, developing deeper networks with more neurons and layers leads to a vanishing gradient problem. Successive studies have introduced several alternative activation functions to reduce the vanishing gradient problem, computational complexity, etc. The most popularly used non-linear activation function for image feature extraction, namely the \emph{Rectified Linear Unit} (ReLU), see \citet{nair2010}. As it represents an almost linear function, it is computationally efficient while allowing for backpropagation. In the dense layers for either classification or regression, non-linear activation functions, namely the Sigmoid, Hyperbolic Tangent(TanH), or similar, can be chosen based on the expected bounded output as a range in contrast to almost linear ReLU outputs. ReLU can be represented as $f(z) = \mathrm{max}(0,z)$ where
\begin{equation}
\mathrm{max}(0,z) =
\begin{cases}
z   &   z \ge 0\;, \\[1mm]
0 & z < 0 \; .
\end{cases}
\end{equation}

\subsection{Pooling Operation}

\begin{figure}[b]
	\centering
	\includegraphics[width=.8\columnwidth]{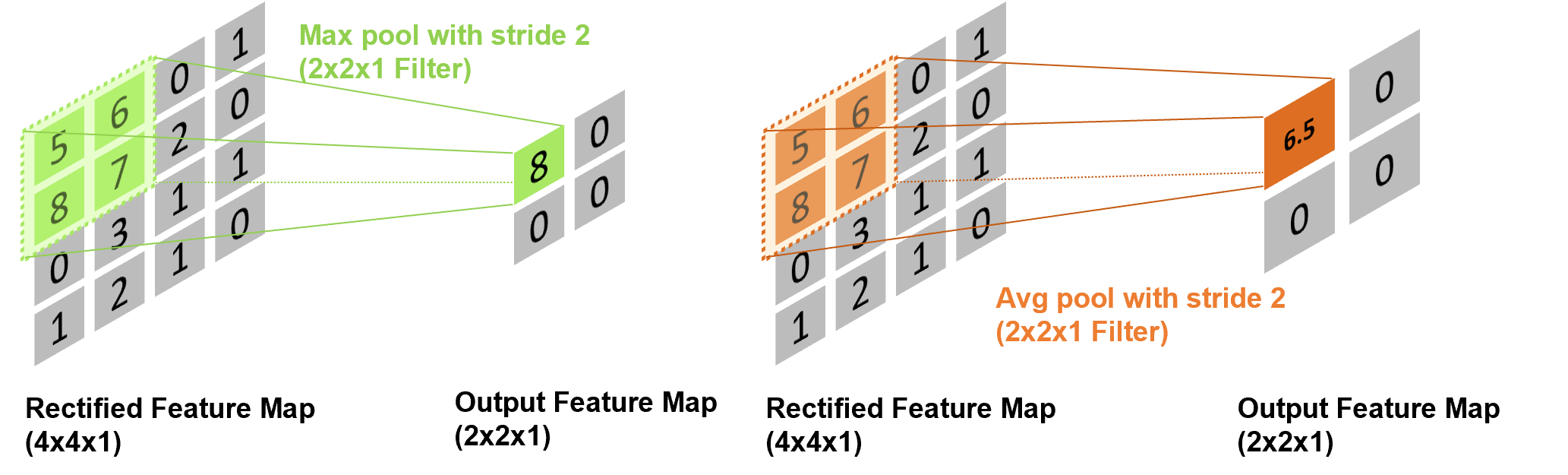}
	\caption{An example of Max and Average Pooling operation performed over feature map generated from convolution operation}
	\label{Average_Max_pooling}
\end{figure}

Another main component of convolutional neural networks is pooling layers, usually placed between two convolutional layers. They reduce the dimensions of the feature map resulting from the convolution operation, thus reducing the learning parameters and computational cost and speeding up the training process. The pooling layer does not have learning parameters. Still, the size of the pooling window and the type of pooling function performed on the pixels in this window are hyper-parameters that need to be tuned during the training process. Normally, this layer's stride length of $S = 2$ will be considered without zero padding. There are many kinds of pooling operations used in CNN models, max pooling and average pooling being the most commonly used ones. An average pooling operation calculates the average value of the pixels in the pooling window. In contrast, in the max pooling operation, the highest pixel value in the pooling window is considered, as shown in Fig \ref{Average_Max_pooling}.

\begin{figure}[htb!]
	\centering
	\includegraphics[width=.55\columnwidth]{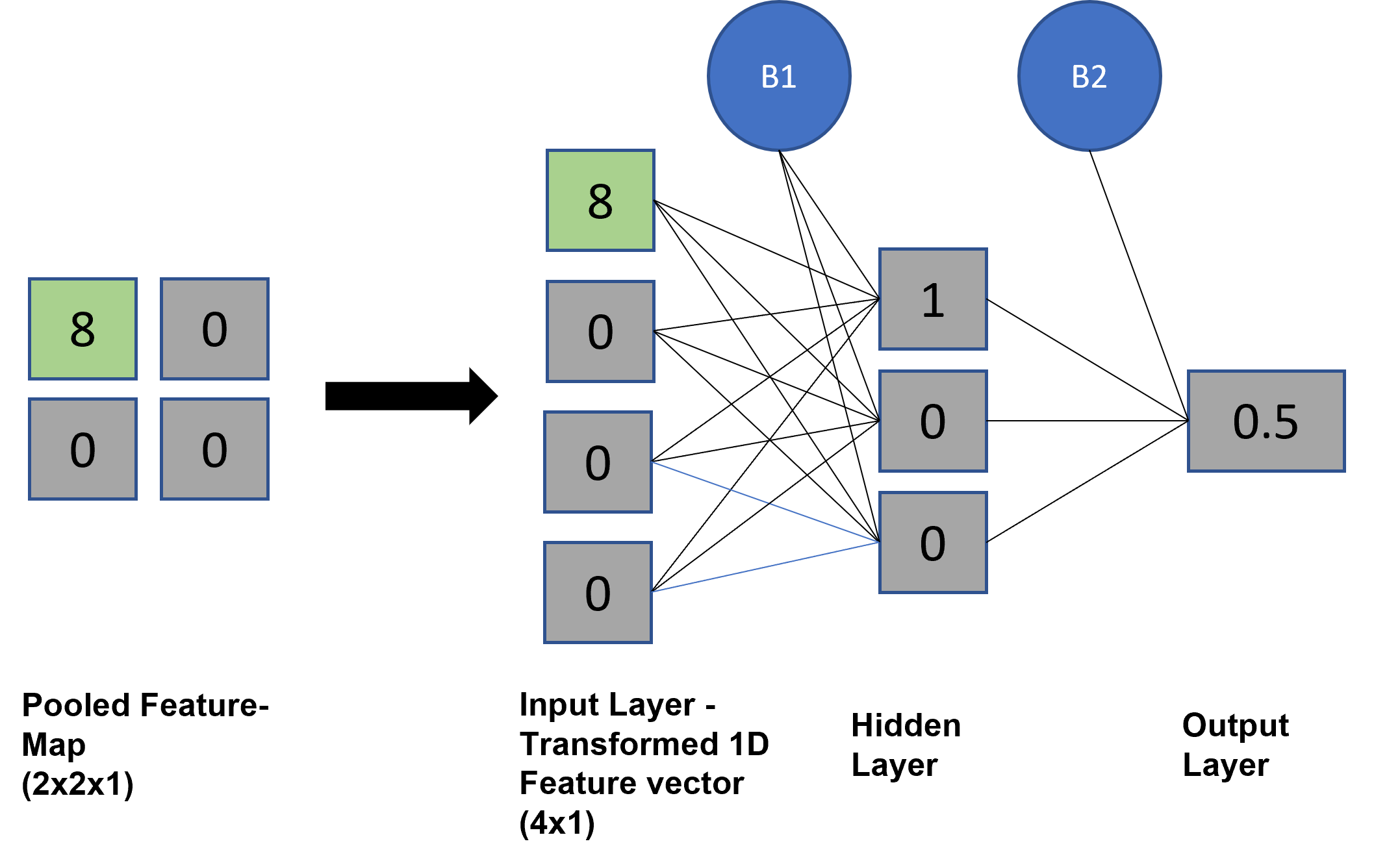}
	\caption{An example of Multi-Layer Perceptron (MLP). B1 and B2 are the biases added to each perceptron cell in the respective hidden and output layers. }
	\label{MLP}
\end{figure}

\subsection{Multi Layer Perceptron (MLP)}
The final component of the CNN is the Multi-Layer Perceptron which consists of a hierarchical connection of layers containing Perceptron cells where information passes from one end to other in making final predictions for matching the ground truth labels. The connections represent the learning
weights that vary during the training process. These adapted weights for each task
represent the learned parameter value and can be detached from the network to perform
another similar task without being retrained. The three important layers collectively define
the MLP architecture, they are \emph{Input layer}, \emph{Hidden Layer} and final \emph{Output Layer}. Fig. \ref{MLP} shows the MLP with connections, where the input layer contains training samples (pooled feature map, in the case of CNN) and is transformed as a 1D feature vector; the hidden layer will follow the input layer with any number of Perceptron cells, where actual processing of information take place by a weighted sum of input and connecting weights with a desired activation function. Output is defined based on the expected number of values matching the ground truth labels.
\subsection{Loss Function}
The optimum parameters for the CNN model are found by optimization of an objective function. The \emph{loss function} refers to the objective function, which is the case for an optimization process using minimization. A loss function measures the error between the model prediction and the ground truth. Thus, it indicates how well the model is performing and should be able to represent this error. That's why different predicting problems require different types of loss functions. For example, in the case of binary classification, binary cross-entropy is a suitable choice. In the case of regression prediction problems, an appropriate loss function could be square error loss or absolute error loss. The loss function is the cost function when applied to the whole data set. A common loss function is the mean squared error loss (MSE). It is the sum of squared differences between predicted and ground truth values, which are expressed by the equation

\begin{equation}
MSE =  \frac{1}{N}\sum_{i=1}^{N} [y_{i} - \widehat{y}_i]^2 \; ,
\end{equation}
where $N$ is the number of training examples, $y_i$ is the ground truth, and $\widehat{y}_i$ are the predicted values.
\\
\\
Once the loss function is defined, during the training process, CNN’s goal is to minimize the error between prediction and ground truth labels in the training samples. Generally, in CNN, there is a sequential process called Forward Propagation which helps in deriving the network output of ground truth label for a given input data with the help of initialized weights. Now, the network output will either be perfectly matched with the ground truth labels or a small error value which has to be minimized by altering the initial weights, which is done with the help of the Back Propagation technique. The backpropagation help in finding the partial derivative of the error function with respect to the weights, thereby minimizing error by subtracting with original weight values. It is clear that repeated forward, and backpropagation in CNN learn the input and gather useful information to activate the desired set of neurons for final results.
\\
\\
During the backpropagation, an optimization algorithm is used to update the weights in a direction close to the global minimum instead of random weight updates. A promising and successful gradient descent algorithm has been used to achieve optimal weight convergence in learning training samples. However, several developments have been introduced to speed up the gradient descent procedure. Notable advancements in gradient descent algorithms are to improve their performance and stability, which includes Batch Gradient Descent, Stochastic Gradient Descent (SGD), and Mini-Batch Gradient Descent. Vanilla or Batch Gradient Descent computes the gradient of a cost function with respect to the parameters considering the entire training samples. In contrast, SGD performs a parameter update only on one training example. Furthermore, Mini-batch gradient descent performs an update for every mini-batch of n training examples. Still, these algorithms have to be optimized better to achieve effective results with less computational cost. The learning rate is an important parameter to bring down the computation of the optimizer algorithm. However, it is quite challenging to define the exact value for the learning rate as a small value leads to a longer convergence time and vice versa. Hence, several optimizers use additional parameters, such as momentum, RMSprop, Adam, etc., for better convergence. Adam optimizer is widely used for image data, with an initialized random search for $N$ number of iterations. In addition, among other techniques, Adam uses SGD as their gradient estimation algorithm to optimize the direction of updating weights.
%
\section{CNN in Magnetostatic Homogenization}
\label{sec4-CNN}
%
This section uses a convolutional neural network model to predict the effective (macroscopic) permeability tensor $\bs \mu^\star$ of artificial microstructures. The proposed model falls under the {\it supervised-learning} category, i.e., the data used for the training and testing processes are labeled. The data generation process, the design of the CNN model architecture, and the results will be discussed in the following subsections.

\begin{figure}[htb!]
	\centering
	\includegraphics[width=0.95\textwidth]{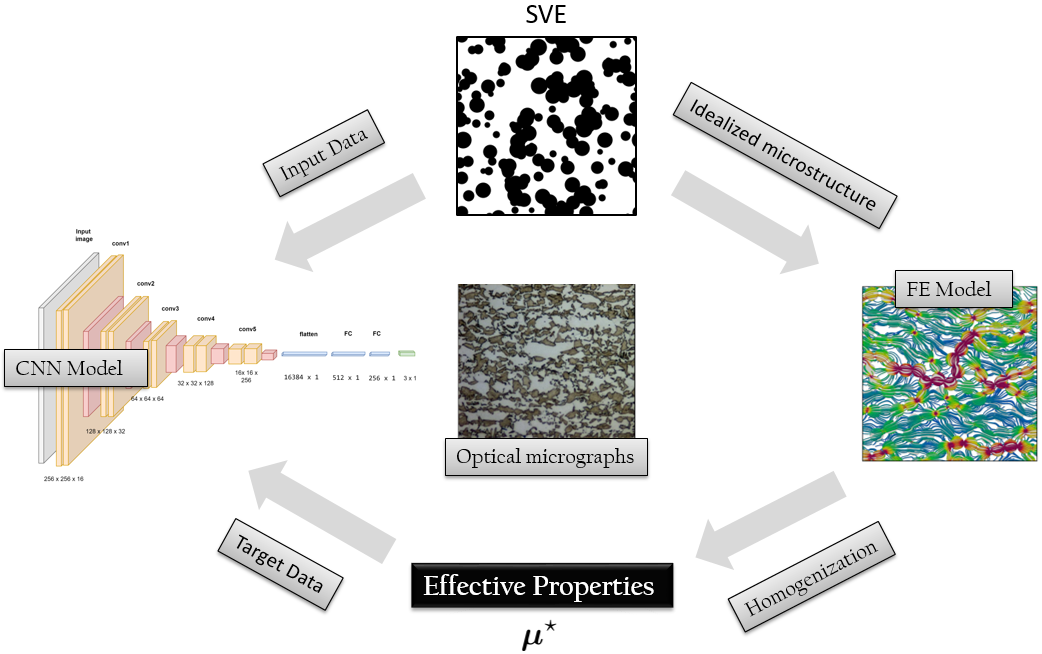}
	\caption{The workflow for the data-generation process. The optical micrograph is that of ferritic-martensitic steel and is taken from Ref.  \cite{ZhouLei2014}. \color{black}}
	\label{data_generation}
\end{figure}

\subsection{Dataset generation}

The macroscopic performance of materials often relies on their microscopic structures, which the naked eye can not see. In recent years, complex microstructures can be captured with high resolution using modern non-destructive imaging techniques such as a micro-computed tomography scan $\mu$-CT or scanning electron microscopes. In this work, the microstructures are synthetically generated.  Effective magnetic permeability determination using finite element-based computational homogenization devising binarized images of these ferritic-martensitic biphasic microstructures
 can be found in, e.g.,  Ref. \cite{SoyarslanetalESAFORM2022}. Artificial periodic and biphasic microstructures with paramagnetic and ferromagnetic constituents possessing relatively high contrast linear magnetic properties are investigated. The workflow of generating the dataset is depicted in Fig \ref{data_generation}.

\begin{figure}[htb!]
	\centering
	\includegraphics[width=\textwidth]{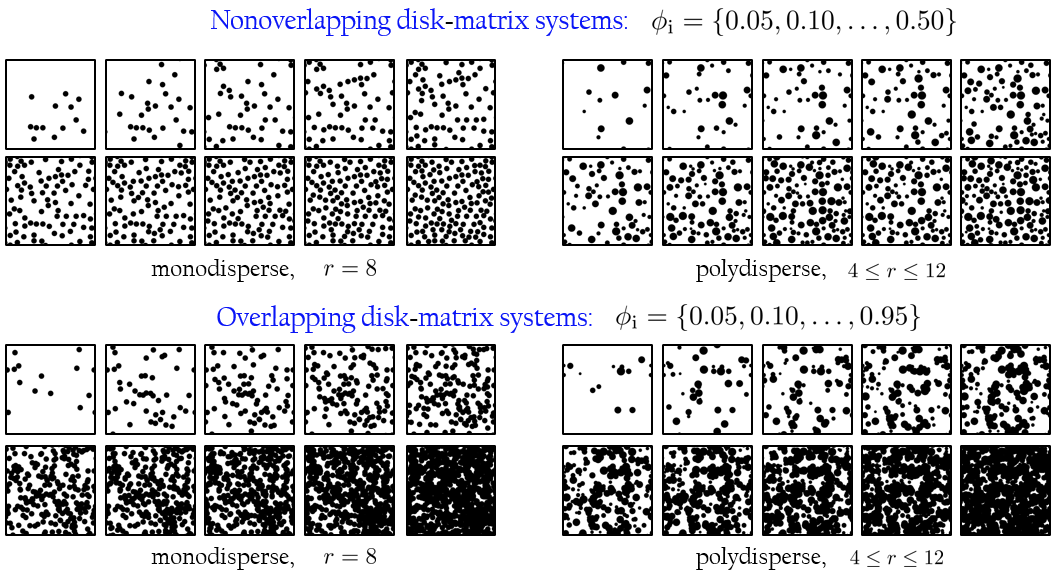}
	\caption{{Data set generation: Images of the idealized microstructure.}}
	\label{idealized_microstructure}
\end{figure}

The generated dataset consists of four groups of nonoverlapping and overlapping, mono- and polydisperse 2D-circular (3D-spherical) disk systems with various volume fractions, which are generated by a random sequential inhibition process, as plotted in Fig. \ref{idealized_microstructure}. Each $\cal{SVE}$ contains a different number of disks (representing the ferromagnetic phase with a larger permeability), resulting in different volume fractions ranging from a minimum of $5\%$ to a maximum of $95\%$.

The resulting geometrical information is then used to produce gray-scale images, where the inclusions are assigned a pixel value of $0$ for the circular inclusions ({ferromagnetic-phase}), {where a pixel value of $0$ is assigned to the circular inclusions ({ferromagnetic-phase})}, and the matrix ({paramagnetic-phase}) is assigned a pixel value of $255$. The dimensions of the 2D images are (256, 256, 1), where the first two numbers represent height and width, and the last number represents the color channels. In this work, gray-scale images are considered the CNN model's input data. To demonstrate the efficiency of the proposed CNN model, we transfer the learning model to a new $\cal{SVE}$ structure and compare the results and computation costs with learning from scratch. For that reason, we started with a simple microstructure; then we tested a complex structure at the end of this work.

\begin{figure}[b]
	\centering
	\includegraphics[width=0.482\textwidth]{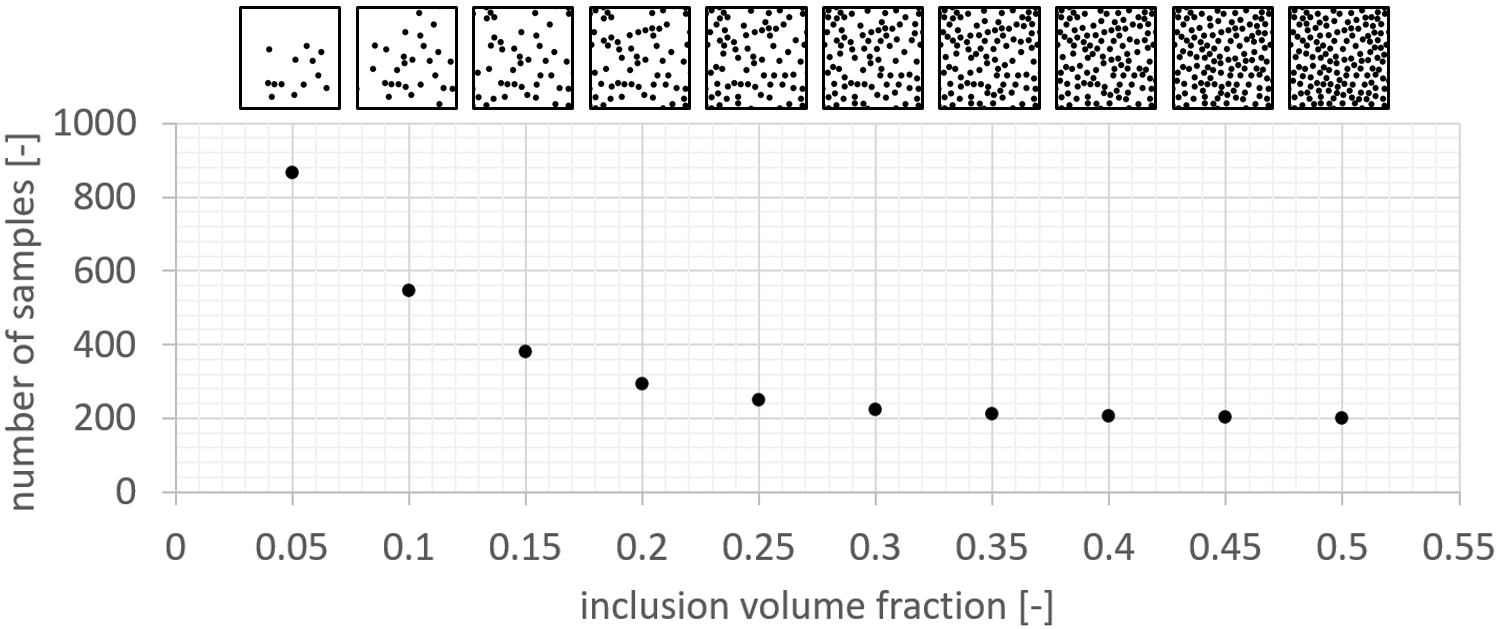} \quad
	\includegraphics[width=0.482\textwidth]{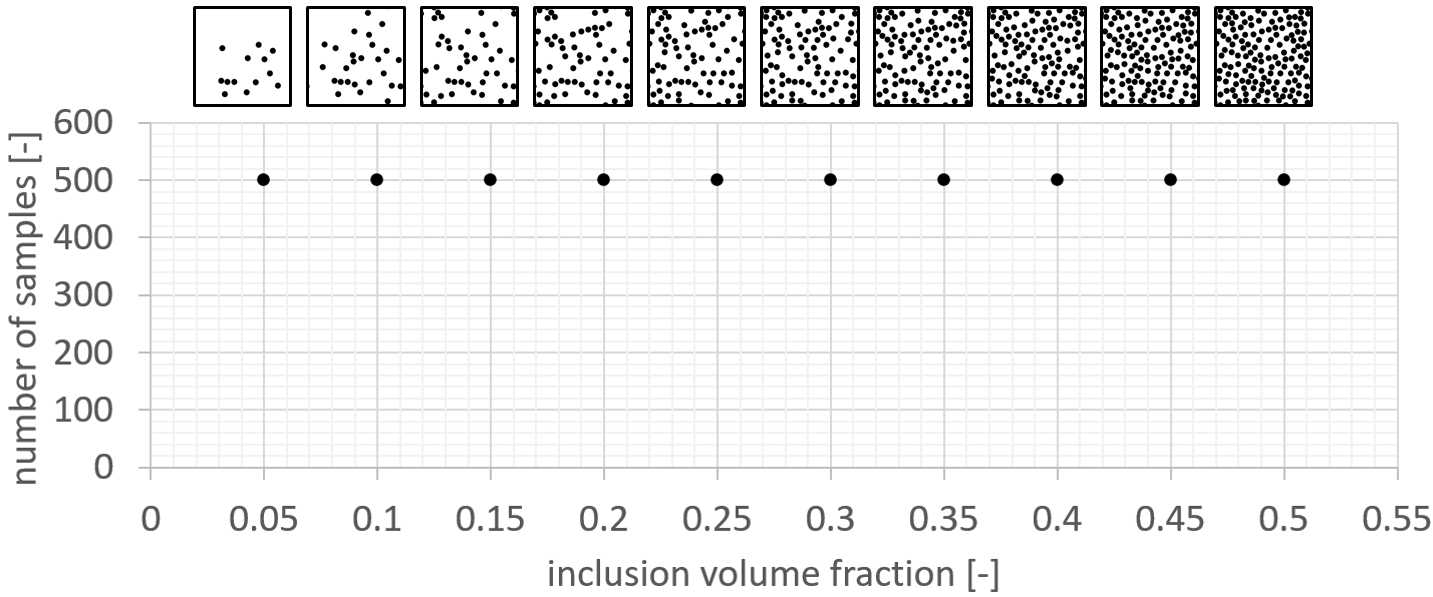}	
	\caption{Data set generation. Sample size distributions. Left: Exponential and Right: Uniform.}
	\label{Sample-distributions}
\end{figure}

\subsection{Calculation of macroscopic permeability tensor}
Acquiring the labels for each one of the generated $\cal{SVE}$ is the second step of generating the dataset, on which the proposed CNN model will be trained and tested. These labels are the components of the macroscopic permeability $\bs \mu^\star$ tensor. The computational multiscale framework introduced in Section \ref{sec2-CNN} is then applied using the {standard finite element method (FEM) to generate the dataset.
Finite element models constitute voxel-based discretizations of the images, which lead to periodic nodal locations, \cite{Soyarslan2022arxiv}. This proves handy in the application of periodic boundary conditions. Evaluation of effective permeability tensor requires two/three loading conditions to be considered in two-/three-dimensional applications. A \textsc{Matlab} script is written to automate the process of pre- and post-processing.} The material properties used in the simulations are summarized in Table \ref{table: Material_properties}.
\begin{table}[h!]
	\centering
	\caption{{Material properties of $\cal{SVE}$ constituents.}}
	\begin{tabular}{||c c ||}
		\hline
		Materials &  Relative magnetic permeability $\mu$  $[-]$   \\ [0.5ex]
		\hline\hline
		Matrix & 1  \\
		\hline
		Inclusion & 250 \\
		\hline
	\end{tabular}
	\label{table: Material_properties}	
\end{table}

Fig. \ref{Sample-distributions} illustrates two sets of sample size distributions for the two-dimensional case.  Our numerical tryouts showed that the exponential distribution (in line with \cite{rao2020three}) shows poor performance in predicting the effective magnetic permeability tensor. Property variations among random microstructure generations at the dilute limit are relatively small for the currently considered high-permeability-inclusion/low-permeability-matrix systems. Thus, we employed a uniform distribution resulting in more satisfactory results close to the FEM reference model. Then, this dataset is divided into three subsets training, validation, and test set. The model uses the training set to learn the parameters (weights, biases). The validation set is used to fine-tune the model's hyper-parameters  (learning rate, number of hidden layers,...) and as an indicator of over-fitting. Finally, the test set gives an unbiased estimation of how well the model generalizes on cases that it has not encountered before.

\begin{figure}[b]
	\centering
	\includegraphics[width=0.8\textwidth]{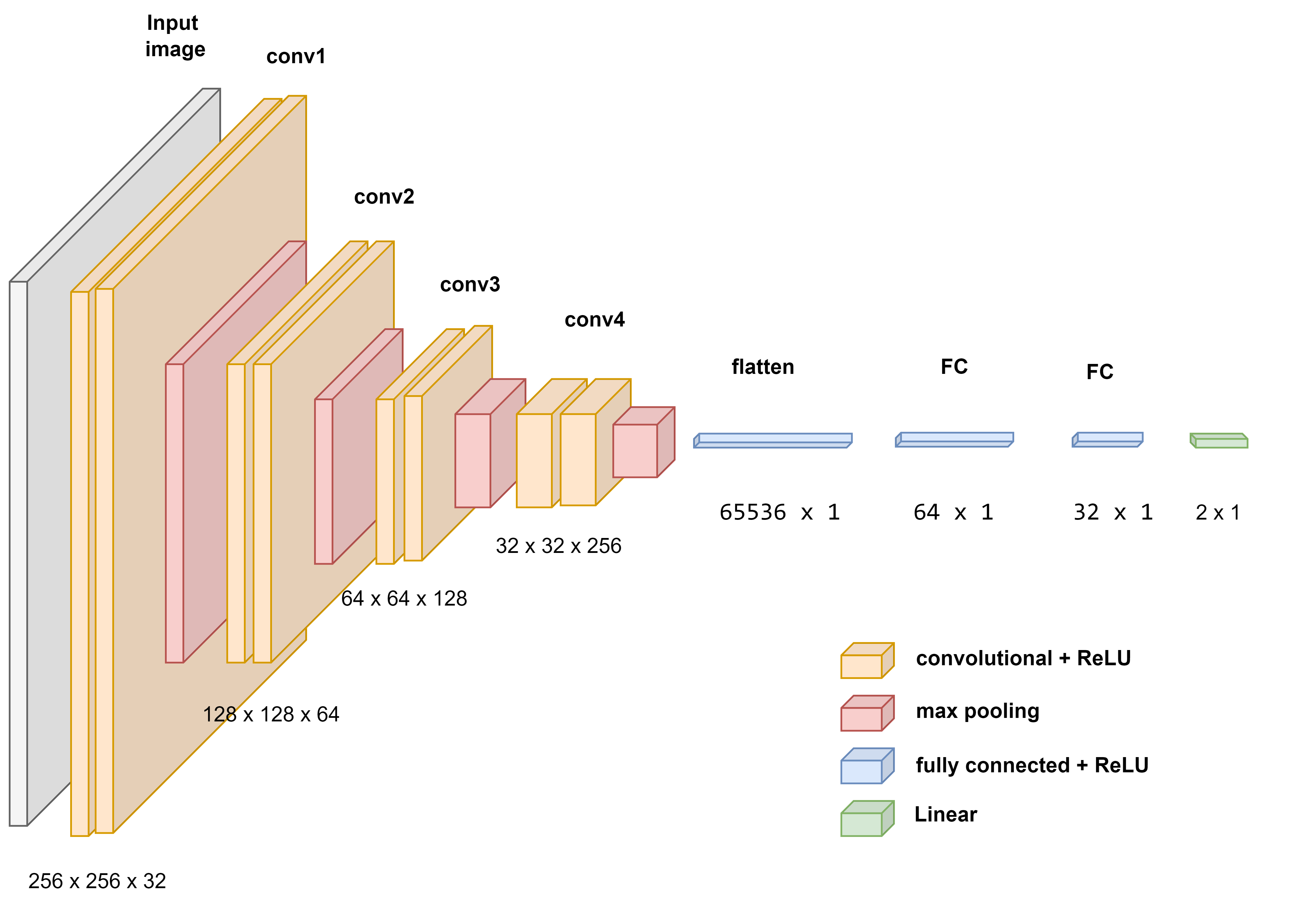}
	\caption{CNN-model architecture for predicting the effective magnetic permeability components.}
	\label{Architecture}
\end{figure}
\subsection{CNN-model architecture}
\begin{figure}[htb!]
	\centering
	\includegraphics[width=0.95\textwidth]{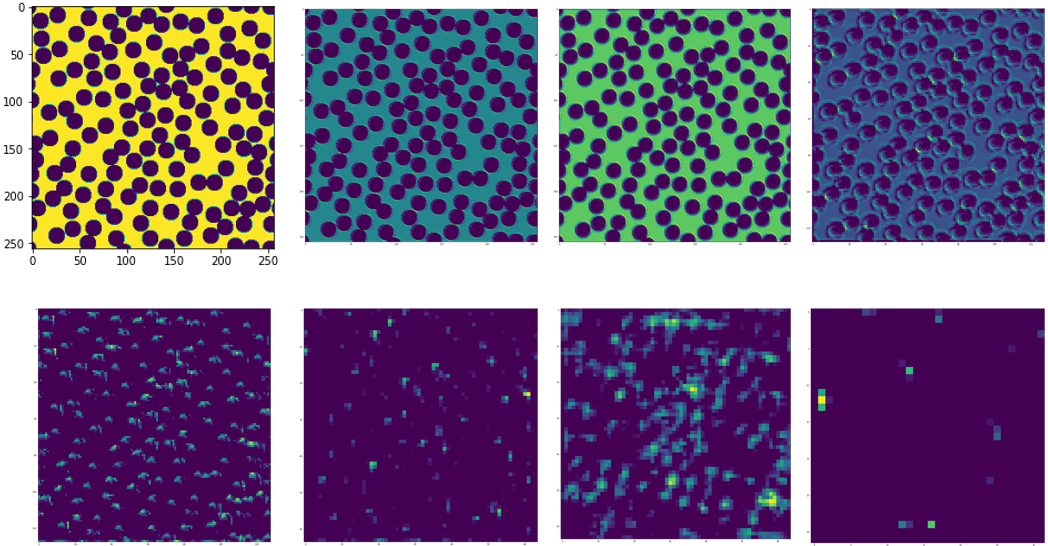}
	\caption{Visualisation of the input image (Two-phases idealized microstructure) and the feature maps learned through the convolutional layers of the CNN-model.}
	\label{Feature_maps}
\end{figure}

The proposed model architecture is sketched in Fig \ref{Architecture}. It is implemented using Keras and Python. The training process was done using Google virtual machines through Google-COLAB, the machine was equipped with a NVIDIA Tesla P100-PCIE-GBU with 16 GB memory. To support the regression task, the mean square error (MSE) between the CNN predictions and the ground truth (results from the finite element simulations) is chosen as a cost function. To have a sufficient feature representation, different filter sizes up to (7,7) are chosen through the convolutional layers for the proposed model. Conv1 and Conv2 layers use (3,3), Conv3 uses (5,5), and the final Conv4 layer uses (7,7) filter kernels, as depicted in Fig \ref{Architecture}. Furthermore, a stride length of $S =1$ and implicit zero padding (same padding) is applied to reduce the effect of a narrower output dimension resulting from the convolution operation. Fig. \ref{Feature_maps} shows the different feature maps through the model. It can be seen that they represent most of the low-level features (shape, edges, and dimension) of the microstructure, but as they go deeper, the learned features seem to be more high-level or task-specific relative feature connections which are hard to be interpreted visually. Next, an MLP network with two hidden layers FC1 and FC2 is used, with 64 and 32 neurons respectively along with a final prediction layer comprising two neurons with a linear activation function. As regularization techniques, $L_2$ regularization with a factor of $0.001$, and early stop with $100$ epochs as a predefined number to stop the training if the validation set error does not decrease, are considered. The activation function used through all the layers is ReLU due to its computational efficiency. The Adam optimizer with a learning rate of $0.0005$ is chosen for training the model while applying a learning rate decay factor of $0.1$ if the validation error does not decrease for $25$ consecutive epochs.

\begin{table}[b]
\centering
	\caption{MAPE and $R^2$ values on the test dataset.}
	\begin{tabular}{||c c c||}
		\hline
		Effective magnetic permeability
 components & $\mu^\star_{11}$ & $\mu^\star_{22}$  \\ [0.5ex]
		\hline\hline
		MAPE & 1.0\% & 1.0\%  \\
		\hline
		$R^2$& 0.994 &0.991 \\
		\hline
	\end{tabular}
	\label{table:Mape and r2}
\end{table}

\begin{table}[b]
	\centering
	\caption{Averaged time needed for computing the magnetic permeability.}
	\begin{tabular}{||c c c  ||}
		\hline
		Computational times  & FEM-Model & CNN-Model \\ [0.5ex]
		\hline\hline
		$100$ different  microstructural images & $40.0~s$ & $0.05~s$ \\
		\hline
	\end{tabular}
	\label{table:time comparison}
\end{table}

\section{Results and Discussions}
\label{sec5-CNN}
%
\subsection{Two Dimensional Applications}
\subsubsection{Biphasic microstructures: Non-overlapping mono-disperse disk-matrix system}

\begin{figure}[htb!]
	\centering
	\includegraphics[width=0.38\textwidth]{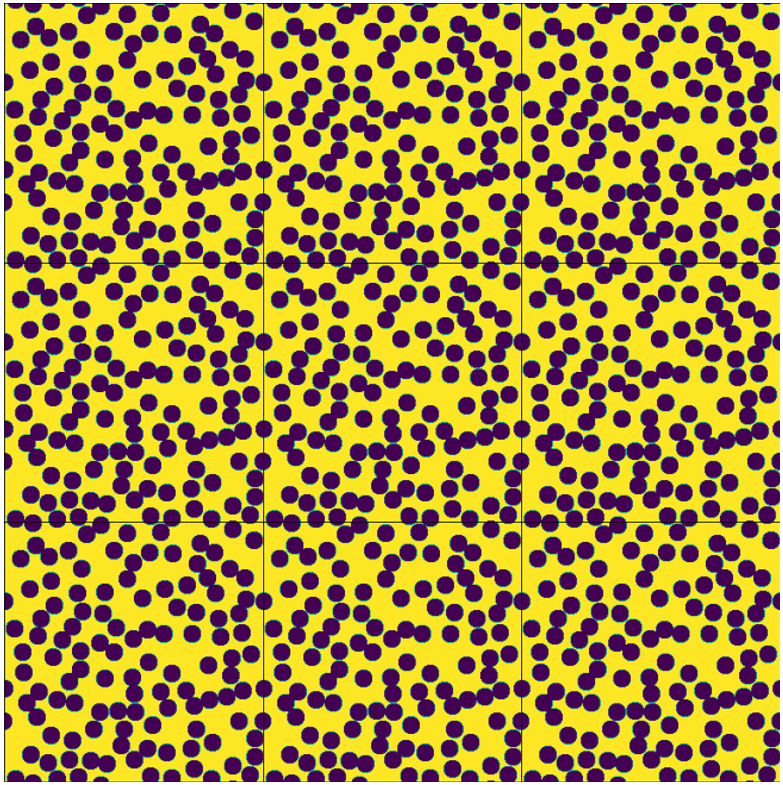}\;
	\includegraphics[width=0.5\textwidth]{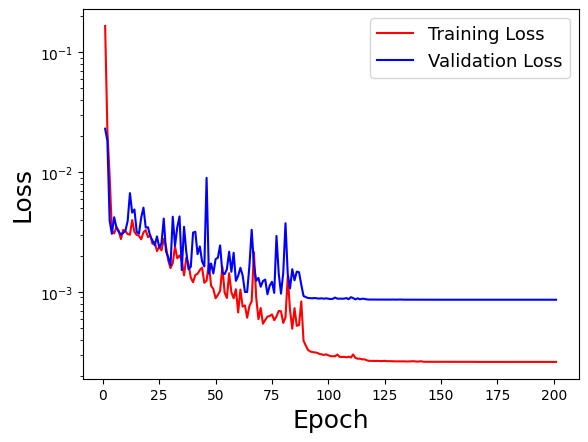}\\[2mm]
\hspace{-80mm}  a) \hspace{70mm} b)
 \\[4mm]
 \includegraphics[width=0.46\textwidth]{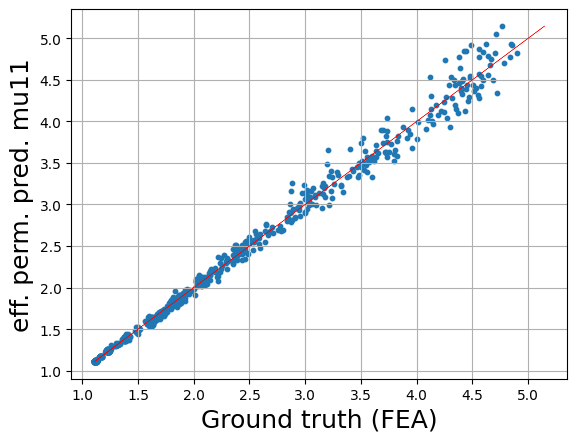}\quad
	\includegraphics[width=0.46\textwidth]{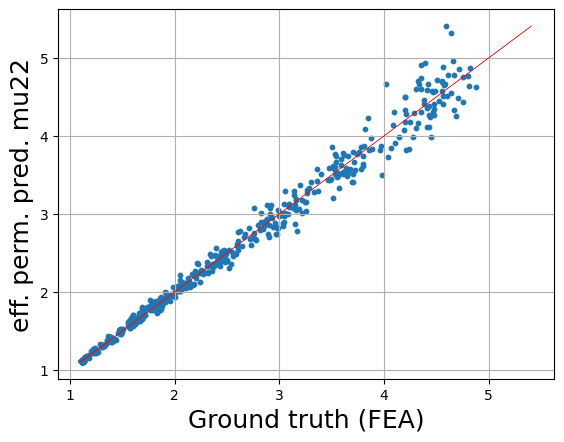} \\
\hspace{-70mm}  c) \hspace{80mm} d)	
 \caption{2D Applications: Non-overlapping mono-disperse disk-matrix system. a) Periodic microstructure generation, b) training and validation losses vs. epochs numbers and c)-d) CNN-predictions vs. ground-truth values of the effective permeability components $\mu^\star_{11}$ and $\mu^\star_{22}$.}
	\label{2D-results-BVP1}
\end{figure}

The first model problem is concerned with predicting the effective magnetic permeability of the biphasic composites introduced in Fig. \ref{idealized_microstructure} with various volume fractions. The details of the CNN-model along with the required data for the FEM-model are described in Section \ref{sec4-CNN}. As a data set generation, we used $5500$ models of a non-overlapping mono-disperse disk-matrix system with a radius of $8$ for the idealized microstructure $\cal{SVE}$. The randomly picked test data sets consist of $825$ $\cal{SVE}$. Those microstructures with two phases are not only stochastic but also periodically distributed, see Fig \ref{2D-results-BVP1}a. Fig \ref{2D-results-BVP1}b shows the MSE on training and validation with respect to the number of epochs. Whereas, the model predictions are shown in Fig \ref{2D-results-BVP1}c-d as a scatter plot, where the red line represents the actual values obtained by finite element simulations (FEM-model). The figure shows the ability of the model to successfully predict the two components of effective magnetic permeability.
\\
\\
The model performance is measured in a quantitative way by the mean absolute percentage error (MAPE) for each component of the permeability tensor. The MAPE is defined as
\begin{equation}
MAPE = \frac{1}{n}\sum_{j=1}^{n}\left|\frac{\widehat{y}_j- y_j}{y_j}\right| \; ,
\end{equation}
where $\widehat{y}_j$ is the predicted value and $y_j$ is  the actual value obtained by FEM simulations. Another method to evaluate the performance of the model is the $R^2$-score, also known as the coefficient of determination, which is a statistical measure that shows how well the CNN-model approximates the actual data. $R^2$-score usually  has a value in range between $0$ and $1$, defined as
\begin{equation}
	R^2 = 1 - \frac{\sum_{j=1}^{n} \big[y_j - \widehat{y}_j \big]^2}{\sum_{j=1}^{n} \big[y_j - \left.\overline{y_j}\right. \big]^2} \; ,
\end{equation}
in terms of the mean value $\left.\overline{y_j}\right.$, where the values closer to $1$ represent a model with better performance. The MAPE and $R^2$-score values for each component of the effective magnetic permeability are given in Table \ref{table:Mape and r2}. The results of the proposed CNN-model are very promising with accurate prediction, which has a good MAPE and coefficient of determination close to $1$.
\\
\\
The key motivation behind employing machine learning in the magnetic field is the high computational efficiency. In this work, the training process of the model took around {$10.0$~min for a total of $300$ epochs}. Once the model is trained, the CNN-model advantages start to kick in, as illustrated in Table \ref{table:time comparison}.

\subsubsection{Transfer learning model vs. training from scratch model}

\begin{table}[b]
\centering
	\caption{Non-overlapping poly-disperse disk-matrix system: Performance comparison between transfer learning model and training from the scratch model: MAPE and $R^2$ values on the test dataset.}
	\begin{tabular}{||c c c||}
		\hline
		Transfer learning model & $\mu^\star_{11}$ & $\mu^\star_{22}$  \\ [0.5ex]
		\hline\hline
		MAPE & 2.0\% & 2.0\%  \\
		\hline
		$R^2$& 0.99 & 0.99 \\
		\hline
	\end{tabular}
\bigskip
	\begin{tabular}{||c c c||}
		\hline
Training from scratch model		 & $\mu^\star_{11}$ & $\mu^\star_{22}$  \\ [0.5ex]
		\hline\hline
		MAPE & 6.0\%  & 5.5\%  \\
		\hline
		$R^2$& 0.97 & 0.97 \\
		\hline
	\end{tabular}
	\label{table:Transfer-learning}
\end{table}
\begin{figure}[htb!]
	\centering
 \caption{2D Applications: Transfer learning model vs. training from the scratch model - Non-overlapping poly-disperse disk-matrix system. a) Data set generation for transfer learning, b) Transfer learning losses (in blue) vs. training from scratch losses (in green), and c)-d) CNN-predictions vs. ground-truth values of the effective permeability components $\mu^\star_{11}$ and $\mu^\star_{22}$.}
	\includegraphics[width=0.52\textwidth]{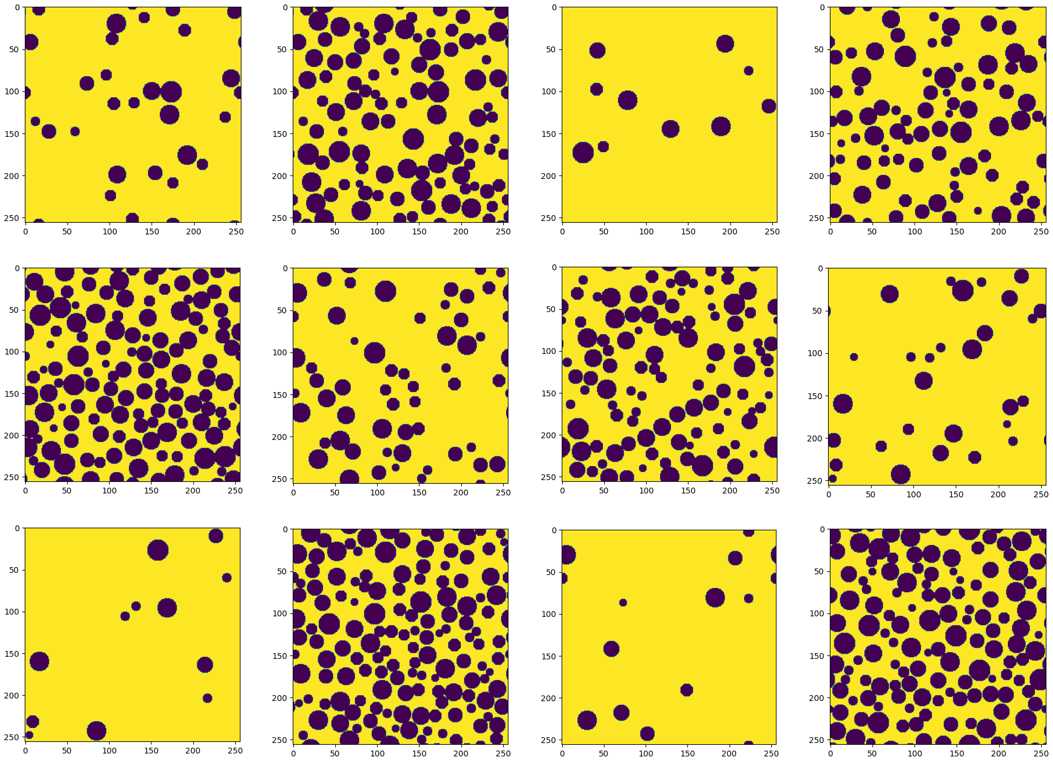}\;
	\includegraphics[width=0.45\textwidth]{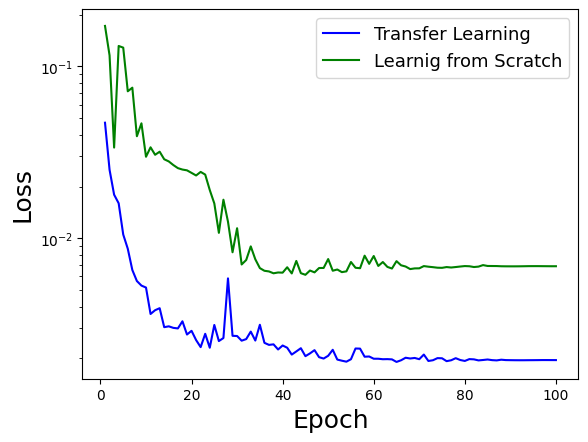}\\
\hspace{-70mm}  a) \hspace{80mm} b)	
 \\[4mm]
 \includegraphics[width=0.48\textwidth]{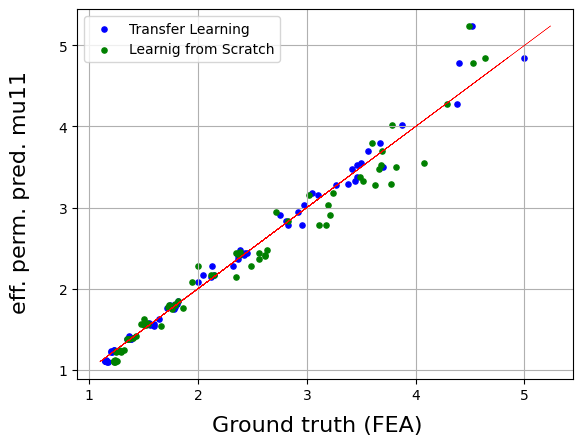}\quad
	\includegraphics[width=0.48\textwidth]{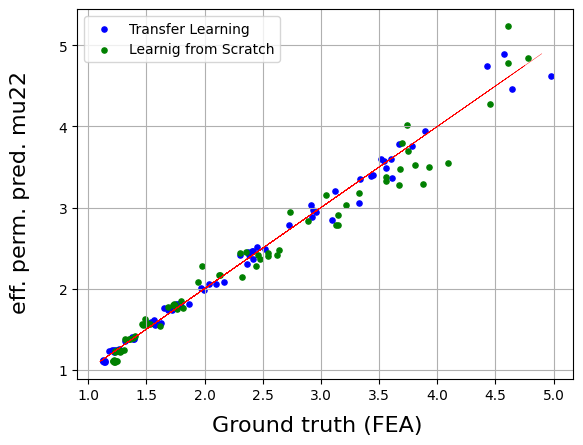} \\
\hspace{-70mm}  c) \hspace{80mm} d)	
	\label{2D-results-BVP2}
\end{figure}

\begin{figure}[htb!]
	\centering
	\includegraphics[width=0.99\textwidth]{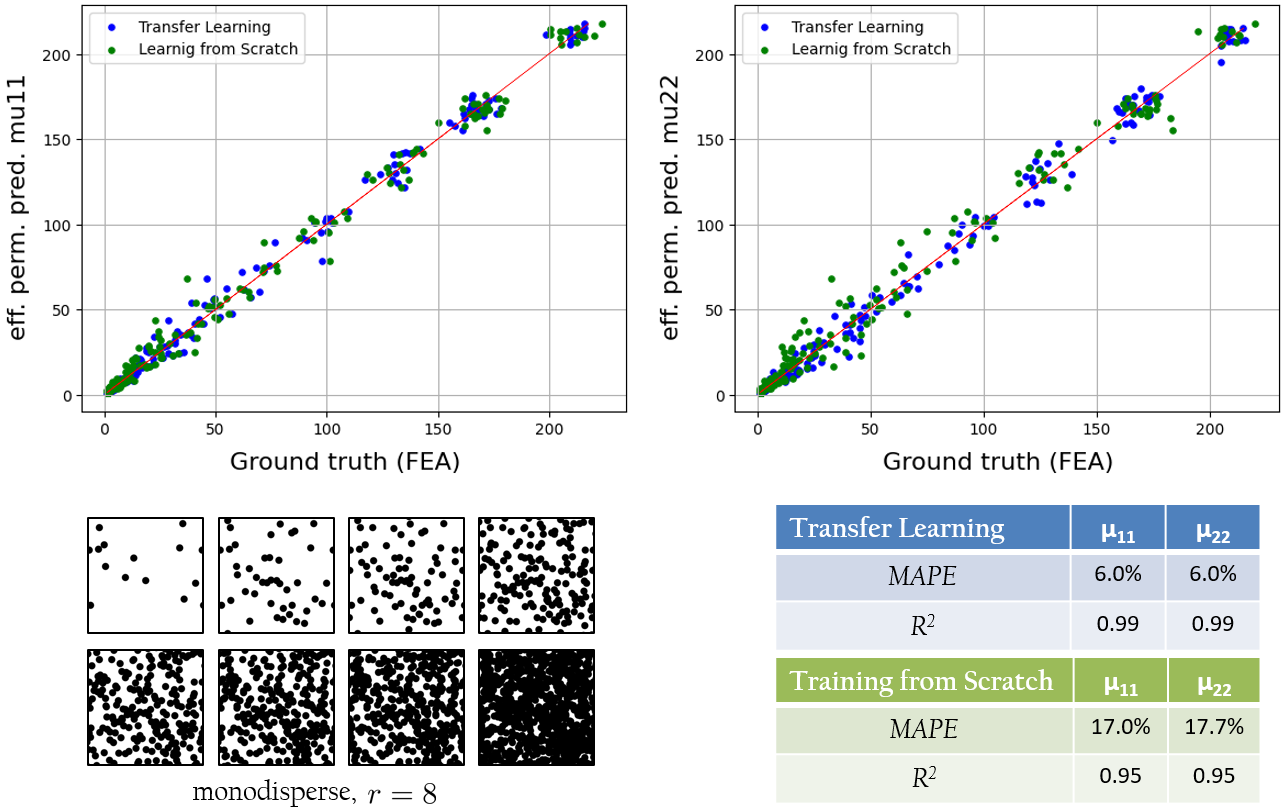}
  \caption{2D Applications: Overlapping mono-disperse disk-matrix system. CNN-predictions vs. ground-truth values of the effective permeability tensor along with the performance comparison between the transfer learning model and training from the scratch model.}
	\label{2D-results-BVP3}
\end{figure}

Next, we demonstrate the capacity of the CNN-model further by transferring the trained model of the previous section to two different cases (data): \emph{Case (i)} Non-overlapping poly-disperse disk-matrix system and \emph{Case (ii)} Overlapping mono-disperse disk-matrix system. The goal here is to predict the effective (macroscopic) permeability tensor $\mu^\star$ using the best-trained model and compare that with training from the scratch model (same procedure as described in the previous example). To this end, we used fewer data (new microstructures), as sketched in Fig \ref{2D-results-BVP2} for {Case (i)} and Fig \ref{2D-results-BVP3} for {Case (ii)}, and only $100$ epochs were considered in the learning process, to demonstrate the CNN transfer-learning efficiency.

As a first comparison, we plot the macroscopic permeability tensor of both models in Figures \ref{2D-results-BVP2}-\ref{2D-results-BVP3} together with the actual values obtained by FEM-model. Hereby, the transfer learning model shows a better performance close to the FEM red-line compared with the training from the scratch model. The good prediction of the transfer learning model required a much less number of epochs (around $20$ epochs) and smaller MSE when compared with the training from the scratch model as illustrated in Fig. \ref{2D-results-BVP2}. Thus, we were able to further accelerate the micro-to-macro simulations of materials with complex micro-structures using the transfer learning approach. Next, we also compare the mean absolute percentage error (MAPE) and the $R^2$-score for both models in Table \ref{table:Transfer-learning} for {Case (i)} and Fig. \ref{2D-results-BVP3} for {Case (ii)}. As expected, the transfer learning model illustrates a better MAPE and coefficient of determination compared with the scratch training model.

\subsection{Extension towards 3D Applications}

\begin{figure}[htb!]
	\centering
	\includegraphics[width=0.18\textwidth]{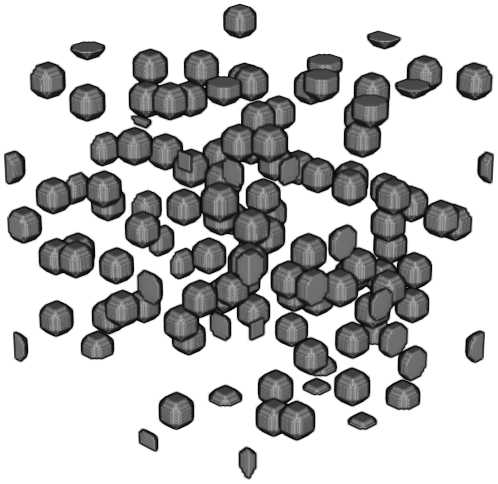}
	\includegraphics[width=0.18\textwidth]{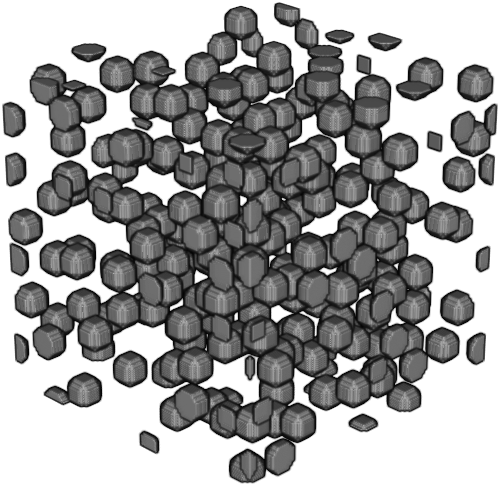}	
 	\includegraphics[width=0.18\textwidth]{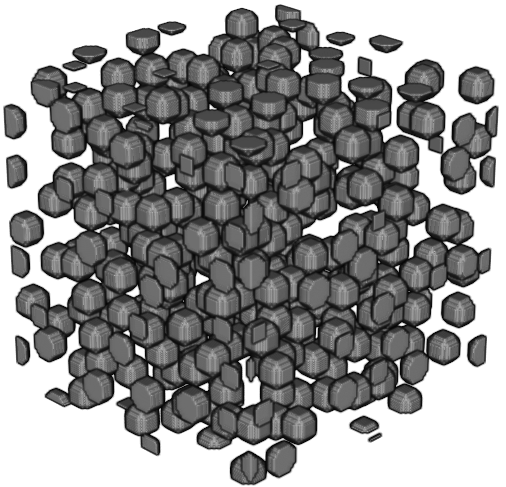}
  \includegraphics[width=0.18\textwidth]{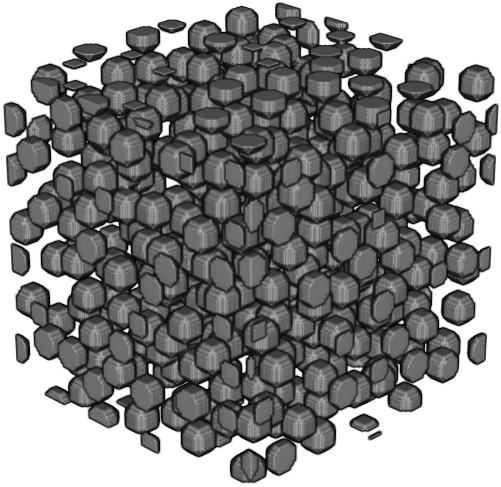}	
		\includegraphics[width=0.18\textwidth]{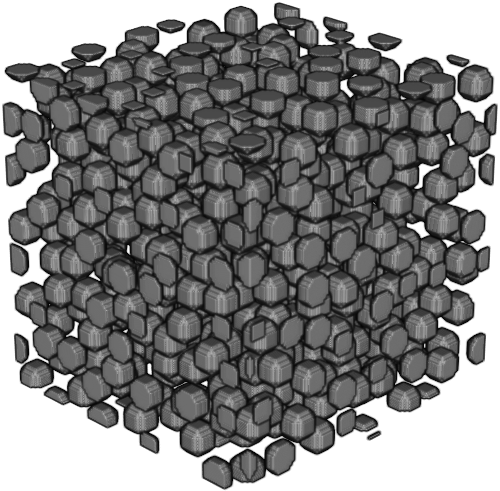}	\\[2mm]
  	\includegraphics[width=0.18\textwidth]{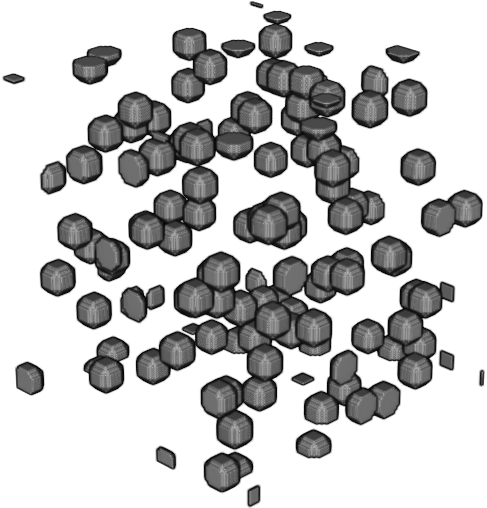}
	\includegraphics[width=0.18\textwidth]{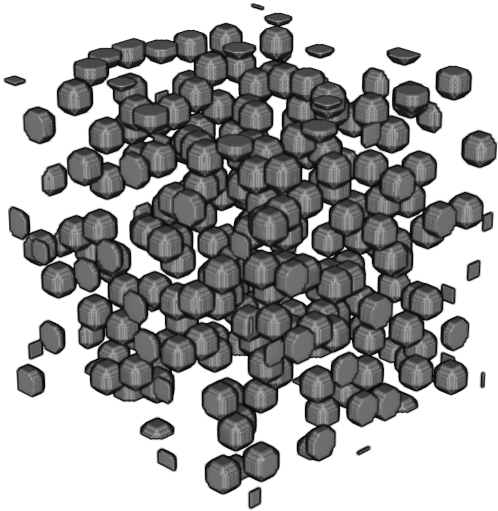}	
 	\includegraphics[width=0.18\textwidth]{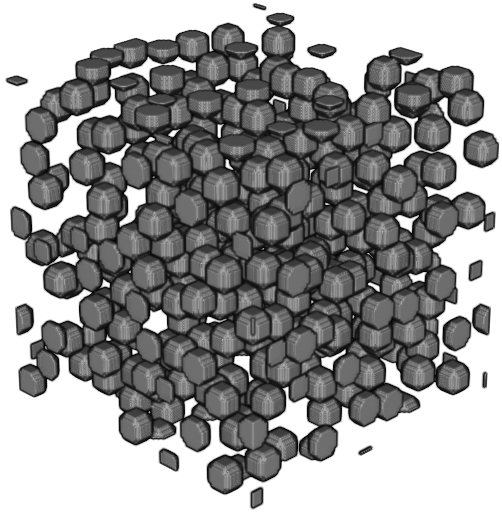}
  \includegraphics[width=0.18\textwidth]{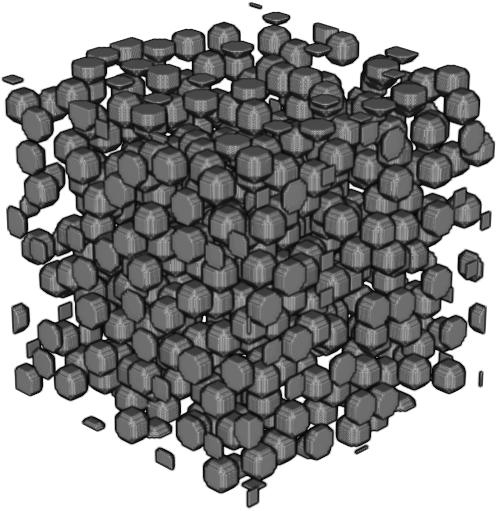}	
		\includegraphics[width=0.18\textwidth]{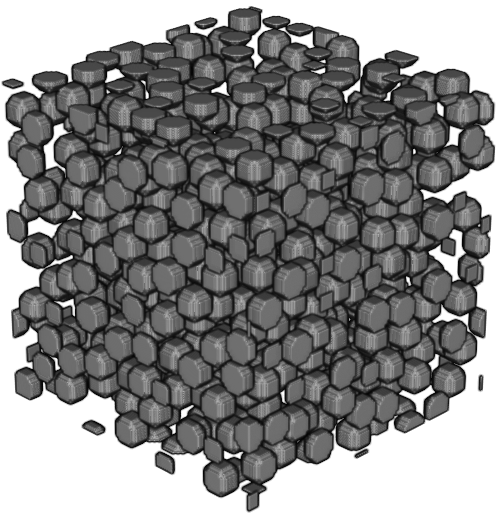}\\[2mm]
		\includegraphics[width=0.18\textwidth]{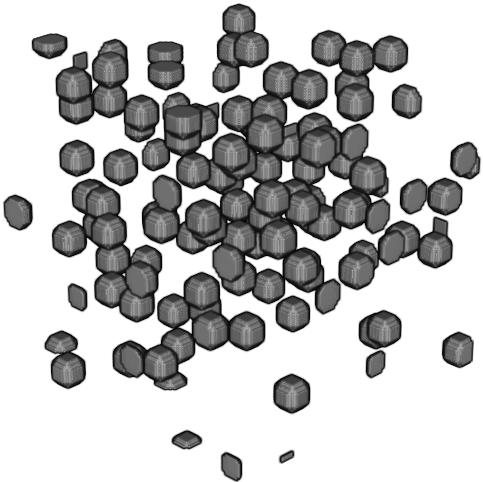}
		\includegraphics[width=0.18\textwidth]{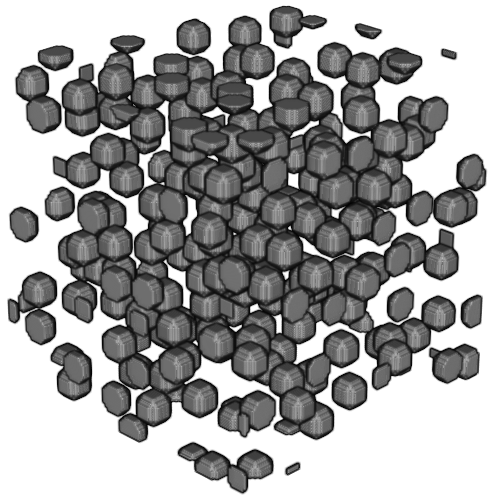}
		\includegraphics[width=0.18\textwidth]{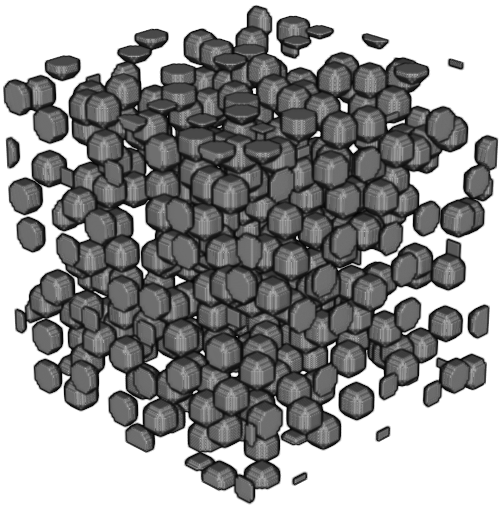}
		\includegraphics[width=0.18\textwidth]{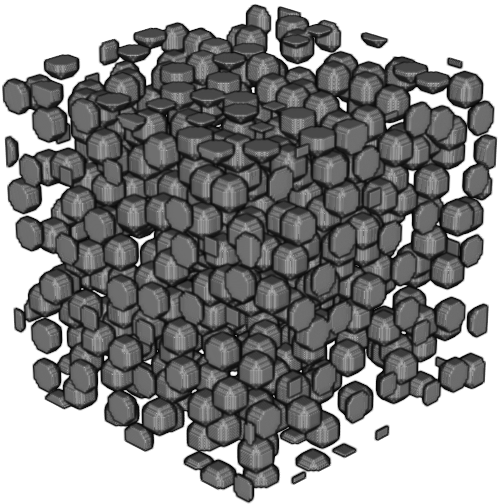}
		\includegraphics[width=0.18\textwidth]{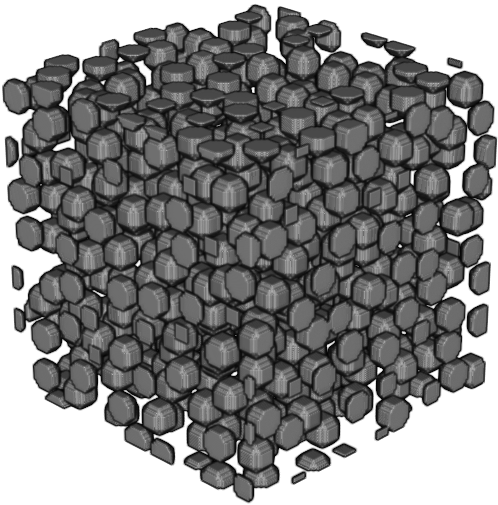}
 \caption{Data set generation. 3D Images of the idealized microstructure.}
	\label{3D-Data}
\end{figure}

\begin{figure}[htb!]
	\centering
	\includegraphics[width=0.48\textwidth]{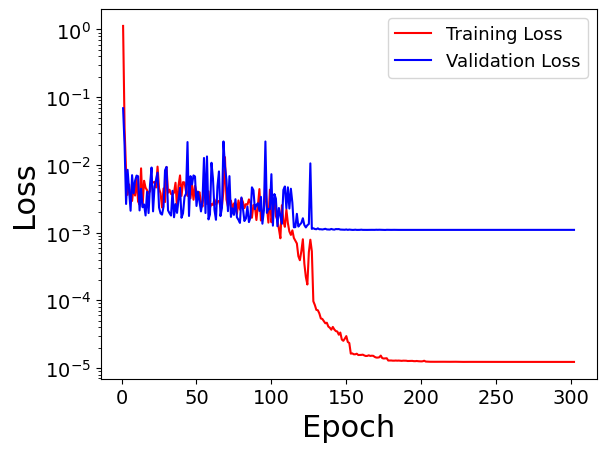}
	\includegraphics[width=0.48\textwidth]{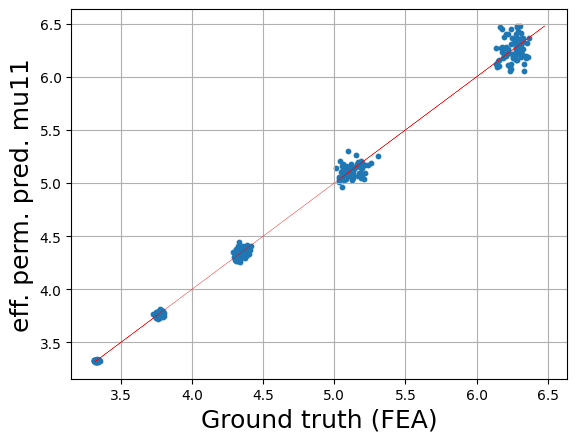} \\
\hspace{-70mm}  a) \hspace{80mm} b)	
 \caption{3D Applications: a) Training and validation losses vs. epochs numbers, and b) CNN-predictions vs. ground-truth values of the effective permeability component $\mu^\star_{11}$.}
	\label{3D-results}
\end{figure}

Finally, we extend the above-introduced two-dimensional investigations toward 3D settings. Similar to the aforementioned study, this example is concerned with predicting the macroscopic effective permeability component $\mu^\star_{11}$. As a data set generation, we used $2500$ models of non-overlapping mono-disperse spheres with a radius of $5$ for the 3D images of the idealized microstructure $\cal{SVE}$, as plotted in Fig \ref{3D-Data}.  To limit the computational cost required for the finite element simulations, a microstructural size of $101\times101\times101$ is selected, similar to the applications presented in Ref.\ \cite{rao2020three}. Within the CNN model, those data are then divided into $70\%$ training data, $15\%$ validation data, and $15\%$ testing data. Hereby, the same architecture used in the 2D case is now extended toward three-dimensional settings.
\\
\\
Fig \ref{3D-results}a shows the error estimations on training and validation with respect to the number of epochs. The model predictions are shown in Fig \ref{3D-results}b as a scatter plot, where the red line represents the values obtained by finite element simulations (FEM-model). The figure shows the ability of the model to predict effective magnetic permeability. Furthermore, the model performance is as well measured by both the mean absolute percentage error (MAPE) and the $R^2$-score
\begin{equation*}
MAPE = 0.8 \% \qquad \mbox{and}  \qquad R^2 = 0.997
\end{equation*}
good predicted results were achieved, similar to the 2D settings.
\\
\\
In this work, the model's training process along with the validation and testing took around {$30.0$~min for a total of $300$ epochs}. Whereas, the computational time required for finite element method-based computations was 15 minutes for a single model. Considering the complete pool to be simulated with 2500 microstructure generations, this ended up with a costly cumulative compute time as compared to the CNN-solution. As in the case of two-dimensional applications, this included the time for pre- and post-processing as well as a job running. To economize time, only one loading case considering loading along $x-$direction is considered. This allowed the construction of only the first column of the effective property tensor.

%
\section{Conclusions}
\label{sec6-CNN}
%
In this contribution, a Convolutional Neural Networks (CNN) model was developed to predict the homogenized  permeability of composites in the case of linear magnetostatics.
In the two- and three-dimensional settings, the input for the CNN model was the images
of artificial periodic and biphasic microstructures in the form of nonoverlapping and overlapping,
mono- and polydisperse circular/spherical disk systems, which are generated by a random sequential inhibition process. These correspond to Statistical Volume Elements (SVE).
The training and testing data for the apparent properties were produced with finite element method-based two-scale asymptotic homogenization.
The proposed CNN-model was chosen after several numerical tests performed on different model architectures.
The results with the CNN model showed high accuracy in predicting the homogenized permeability and a significant decrease in computation time.

\smallskip
\smallskip

{\bf Acknowledgment.}
{Fadi Aldakheel} (FA) gratefully acknowledges support for this research by the ``German Research Foundation'' (DFG) in the {\sc Priority Program SPP 2020} within its second funding phase.
\clearpage
\bibliography{bibliography}
\end{document}